\documentclass[eqsecnum,twocolumn,aps,epsf,superscriptaddress]{revtex4}
\usepackage{graphicx}
\usepackage{color}
\begin{document}
\draft

\title{Novel Electronic State and Superconductivity in the Electron-Doped High-$T_{\rm c}$ T'-Superconductors}

\author{T. Adachi}
\thanks{Corresponding author: t-adachi@sophia.ac.jp}
\affiliation{Department of Engineering and Applied Sciences, Sophia University, 7-1 Kioi-cho, Chiyoda-ku, Tokyo 102-8554, Japan}

\author{T. Kawamata}
\author{Y. Koike}
\affiliation{Department of Applied Physics, Tohoku University, 6-6-05 Aoba, Aramaki, Aoba-ku, Sendai 980-8579, Japan}

\begin{abstract}
In this review article, we show our recent results relating to the undoped (Ce-free) superconductivity in the electron-doped high-$T_{\rm c}$ cuprates with the so-called T' structure.
For an introduction, we briefly mention the characteristics of the electron-doped T'-cuprates, including the reduction annealing, conventional phase diagram and undoped superconductivity.
Then, our transport and magnetic results and results relating to the superconducting pairing symmetry of the undoped and underdoped T'-cuprates are shown.
Collaborating spectroscopic and nuclear magnetic resonance results are also shown briefly.
It has been found that, through the reduction annealing, a strongly localized state of carriers accompanied by an antiferromagnetic pseudogap in the as-grown samples changes to a metallic and superconducting state with a short-range magnetic order in the reduced superconducting samples. 
The formation of the short-range magnetic order due to a very small amount of excess oxygen in the reduced superconducting samples suggests that the T'-cuprates exhibiting the undoped superconductivity in the parent compounds are regarded as strongly correlated electron systems, as well as the hole-doped high-$T_{\rm c}$ cuprates. 
We show our proposed electronic structure model to understand the undoped superconductivity.
Finally, unsolved future issues of the T'-cuprates are discussed.

\end{abstract}
\vspace*{2em}
\pacs{PACS numbers: }
\maketitle
\newpage

\section{Introduction}
\begin{figure*}[tbp]
\centering
\includegraphics[width=0.8\linewidth]{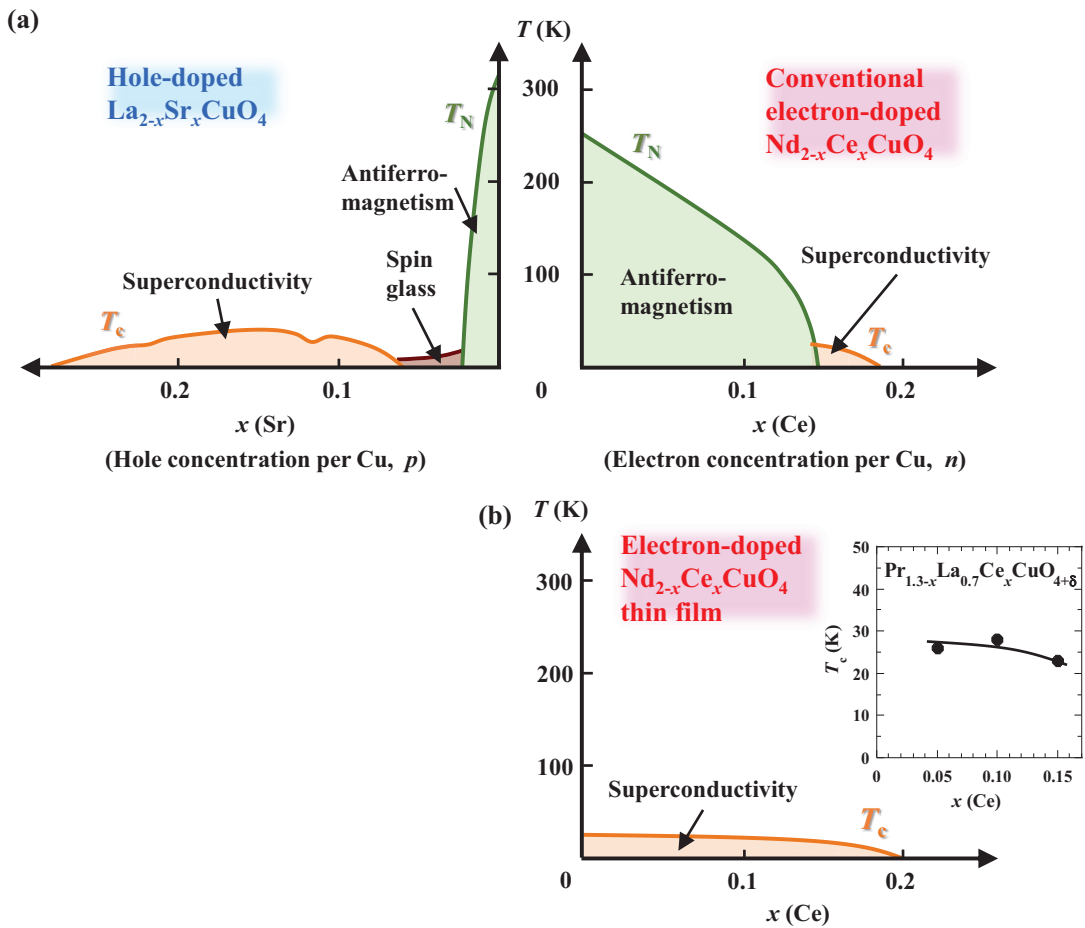} 
\caption{Phase diagrams of (\textbf{a}) the hole-doped and conventional electron-doped high-$T_{\rm c}$ cuprates and (\textbf{b}) the electron-doped cuprates newly proposed in thin films. The inset of ({b}) is the Ce concentration dependence of $T_{\rm c}$ in single crystals of electron-doped Pr$_{1.3-x}$La$_{0.7}$Ce$_x$CuO$_{4+\delta}$.}
 \label{figure1}
%\end{center}
\end{figure*}

In the past three decades, cuprates exhibiting high-temperature superconductivity have been intensively studied.
The superconducting (SC) transition temperature $T_{\rm c}$ $\sim 134$ K \cite{Hg-1223} in the Hg-1223 cuprate is the highest among all SC materials at ambient pressure.
Although new high-$T_{\rm c}$ superconductors of iron arsenides with the maximum $T_{\rm c}$ of $\sim 55$ K \cite{Sm-1111} and of sulfur hydrides with $T_{\rm c}$ $\sim 203$ K at 200 GPa \cite{H3S} have been observed, the cuprates have been standing for potential materials of both exotic physics and future applications. 
All series of high-$T_{\rm c}$ cuprates, La-214, Y-123, Bi-2212, etc., have the parent compounds, which are antiferromagnetic (AF) Mott insulators.
As shown in the phase diagram of Figure \ref{figure1}a, in the hole-doped cuprates, doping of hole carriers into the parent compound destroys an AF order quickly, and then, the superconductivity appears.
With increasing hole doping, $T_{\rm c}$ is raised in the underdoped regime and exhibits the maximum in the optimally doped regime.
With further doping, $T_{\rm c}$ turns into a decrease in the overdoped regime, and finally, the superconductivity disappears in the heavily-overdoped regime where the system becomes a metal.
Comprehensive neutron scattering~\cite{Birgeneau-JPSJ} and muon spin relaxation ($\mu$SR)~\cite{Watanabe,Adachi-PRB,Risdi-LSCO,Adachi-PRBopt,Koike-physC} experiments have uncovered AF fluctuations in the SC doping regime, suggesting that AF fluctuations are a glue to form SC electron pairs.
In the non-SC heavily-overdoped regime, recent theories~\cite{Kopp,Barbiellini,Jia} and experiments~\cite{Sonier,Kurashima} have suggested the existence of ferromagnetic fluctuations of itinerant carriers.

Electron-doped high-$T_{\rm c}$ cuprates were first discovered by Tokura et al. in 1989~\cite{Tokura}.
Pronounced features different from those in the hole-doped cuprates are not only the electron carrier, but also the reduction annealing, which is absolutely essential to obtain the superconductivity in the electron-doped cuprates.
The phase diagram depicted by the {{normally}} reduced samples is shown in Figure \ref{figure1}a, which roughly resembles that of the hole-doped cuprates. 
The doping of electrons into the parent compounds weakens the AF order, but the AF order is maintained around the optimally doped regime.
The superconductivity with the maximum $T_{\rm c}$ appears around the optimally doped regime and is monotonically suppressed with over-doping. 
Since AF fluctuations were observed around the optimally doped regime~\cite{Yamada-PRL}, the electron pairing mediated by AF fluctuations has been believed to be the case also in the electron-doped cuprates.
In 2005, a surprising result was obtained in thin films of the electron-doped cuprates with the Nd$_2$CuO$_4$-type structure (the so-called T' structure), in which the superconductivity appears without electron doping~\cite{Tsukada}. 
Following observations of the {{undoped}} (Ce-free) superconductivity in the parent compounds~\cite{Asai,Takamatsu} and the suggestion of a new phase diagram shown in Figure \ref{figure1}b~\cite{Matsumoto} opened a new era of research in the high-$T_{\rm c}$ superconductivity. %Is the italics necessary? 
Yet to date, the mechanism of the undoped superconductivity is unclear.

In this review article, we show our recent results relating to the undoped superconductivity in the T'-cuprates~\cite{Adachi-JPSJ,Adachi-JPSJ2,Ohashi}.
First, we present the characteristics of the electron-doped T'-cuprates including the reduction annealing, conventional phase diagram and undoped superconductivity.
The next three sections consist of our transport and magnetic results and results relating to the SC pairing symmetry.
Collaborating spectroscopic and nuclear magnetic resonance (NMR) results are also shown briefly.
Then, we discuss our proposed electronic structure model to understand the undoped superconductivity.
In the final section, unsolved future issues of the T'-cuprates are discussed.
Note that there are already excellent and complete reviews on the materials and physics focusing on the conventional electron-doped cuprates~\cite{Armitage,Fournier}.

%%%%%%%%%%%%%%%%%%%%%%%%%%%%%%%%%%%%%%%%%%
\section{Electron-Doped T'-Cuprate}

Electron-doped T'-cuprates were discovered in Nd$_{2-x}$Ce$_x$CuO$_4$ ~\cite{Tokura}.
This class of materials are expressed as $RE_2$CuO$_4$ ($RE$ = Pr, Nd, Sm, Eu) with the T' structure.
The electron doping is achieved by the substitution of the tetravalent Ce$^{4+}$ for the trivalent $RE^{3+}$.
The T' structure consists of the CuO$_2$ plane and fluorite-type blocking layer.
Because as-grown T'-samples are insulating and AF, the reduction annealing for the as-grown T'-cuprates is indispensable to obtain superconductivity.
One of the unsolved puzzles for the T'-cuprates is the exact mechanism of the reduction process.
To date, there are mainly two candidate mechanisms for this puzzle.
One is the removal of excess oxygen from the as-grown sample, which is mostly believed from the early stage of the research of the T'-cuprates.
This idea originated from early neutron-diffraction experiments in Nd$_{2-x}$Ce$_x$CuO$_4$~\cite{Radaelli,Schultz}.
Because of the size mismatch between the CuO$_2$ plane and blocking layer, excess oxygen tends to be included right above Cu, resulting in the shrinkage of the blocking layer in the $ab$-plane.
Since the excess oxygen is understood to induce disorder of the electrostatic potential in the CuO$_2$ plane, electron pairs are destroyed, and the system shows no SC behaviors~\cite{Xu}. 
Therefore, the removal of excess oxygen from the as-grown sample is crucial for the appearance of superconductivity and the investigation of intrinsic properties of the T'-cuprates. 
The other is the Cu-vacancy scenario proposed by neutron and X-ray diffraction studies in Nd$_{2-x}$Ce$_x$CuO$_4$~\cite{Mang} and Pr$_{1-x}$LaCe$_x$CuO$_4$~\cite{Kang}.
The diffraction experiments found that (i) the occupancy at the Cu site is imperfect (perfect) in the as-grown (reduced) samples and (ii) the so-called secondary phase of $RE_2$O$_3$ appears (disappears) through the reduction (oxidation) annealing, both of which are reversible processes through the reduction (oxidation) annealing.
Therefore, the role of the reduction annealing is to fill up Cu deficiencies in the as-grown samples where Cu deficiencies are regarded as causing the pair breaking.
The secondary phase was precisely investigated in Nd$_{2-x}$Ce$_x$CuO$_4$~\cite{Kimura-JPSJ}.

The phase diagram of T'-cuprates was first obtained by $\mu$SR in Nd$_{2-x}$Ce$_x$CuO$_4$ (Figure \ref{figure1}a)~\cite{Luke} and later in Pr$_{1-x}$LaCe$_x$CuO$_4$~\cite{Fujita-PD}.
The AF phase expands up to the optimally doped regime, and the optimal superconductivity suddenly sets in at the same time of the suppression of the AF order.
In detail, both the superconductivity and AF order coexist around the boundary of two phases. 
With over-doping, $T_{\rm c}$ monotonically decreases and disappears around the region of the solubility limit of Ce into the T'-structure.
Neutron-scattering experiments in Nd$_{2-x}$Ce$_x$CuO$_4$ around the optimally doped regime of $x=0.15$ revealed commensurate AF fluctuations~\cite{Yamada-PRL}.
This is in sharp contrast to incommensurate AF fluctuations~\cite{Birgeneau-JPSJ} or stripe fluctuations~\cite{Tranquada} observed in the hole-doped cuprates.
Following neutron-scattering experiments in Pr$_{1-x}$LaCe$_x$CuO$_4$~\cite{Fujita-PRL} uncovered that commensurate AF fluctuations existed also in the overdoped regime and that both the spin stiffness and relaxation rate of AF fluctuations decreased toward the end point of superconductivity in the heavily overdoped regime.
These suggest the intimate relation between the superconductivity and AF fluctuations in the T'-cuprates, as well as in the hole-doped cuprates.

In 1995, Brinkmann et al. reported pioneering results, which later led to the discovery of undoped (Ce-free) superconductivity~\cite{Brinkmann}.
They tried to expand the SC regime deeply into the AF underdoped regime in the phase diagram by improving the reduction process.
A thin single crystal of Pr$_{2-x}$Ce$_x$CuO$_4$ was sandwiched by polycrystalline pellets of the same component to protect the surface of the crystal and annealed at a higher temperature of 1080 $^{\circ}$C for a longer time than in the conventional annealing.
They observed the superconductivity down to $x=0.04$ in the underdoped regime. 
Moreover, $T_{\rm c}$ gradually increased with decreasing doping from the optimally doped regime.
They suggested that the superconductivity in underdoped Pr$_{2-x}$Ce$_x$CuO$_4$ was a product of the appropriate reduction of excess oxygen.
Then, one would think of lower doping regime than $x=0.04$, namely whether Mott insulators like the hole-doped cuprates or superconductivity down to the parent compound. 

Ten years later, a surprising result was obtained by Tsukada et al. in T' thin films of Ce-free La$_{2-y}RE_y$CuO$_4$ ($RE$ = Y, Lu, Sm, Eu, Gd, Tb) in which the superconductivity appears without Ce substitution (electron doping)~\cite{Tsukada}.
Subsequent systematic investigation by the same group using Nd$_{2-x}$Ce$_x$CuO$_4$ thin films uncovered that $T_{\rm c}$ of $\sim 28$ K in the parent compound of $x=0$ decreased monotonically with increasing $x$ and disappeared at $x \sim 0.20$ (Figure \ref{figure1}b)~\cite{Matsumoto}.
These findings were suggested to be due to the removal of excess oxygen from the as-grown thin films more effectively than that ever reported. 
Their unique procedure of the reduction annealing called two-step annealing was mentioned in the literature~\cite{Krockenberger-SciRep}.
At the first step, the as-grown thin films were annealed at a high temperature in an intermediate oxygen pressure of $\sim 10^2$ Pa.
In this process, they insisted based on the increasing $ab$-plane electrical resistivity $\rho_{\rm ab}$ and unchanged $c$-axis length that oxygen in the CuO$_2$ plane was mainly removed, and excess oxygen was not. 
At the second step, the thin films were further annealed at a low temperature in a low oxygen pressure of $\sim 10^{-5}$ Pa.
In this process, they insisted based on the decreasing $\rho_{\rm ab}$ and $c$-axis length that excess oxygen moved to the oxygen-defect site in the CuO$_2$ plane, and the superconductivity was realized. 
In fact, optical studies of the SC thin film of Pr$_2$CuO$_x$ with $x \simeq 4$ have revealed a Drude-like peak centered at zero frequency, suggesting a metallic state of the SC parent compound~\cite{Chanda}.
These results give us the following messages: (i) the newly-proposed phase diagram without the AF phase is seemingly different from the conventional one; and (ii) the AF Mott insulating state is not a starting point of the superconductivity in the T'-cuprates.
The superconductivity in the parent compounds of the T'-cuprates was also confirmed using polycrystalline samples of Ce-free La$_{2-y}$Sm$_y$CuO$_4$~\cite{Asai} and Ce-free La$_{1.8}$Eu$_{0.2}$CuO$_4$ ~\cite{Takamatsu}, while it has not yet been confirmed in single crystals of the parent compound.

From the theoretical viewpoint, the undoped (Ce-free) superconductivity in the parent compounds might be explained by the early local density energy band calculation in Nd$_{2-x}$Ce$_x$CuO$_4$~\cite{Massidda} where the system was a simple band metal without electron correlation. 
On the assumption of the moderate electron correlation, the half-filled-band Hubbard model consisting of doublons and holons predicted the superconductivity in the parent compound~\cite{Yokoyama}.
Under the strong electron correlation generating the Mott--Hubbard gap between the upper Hubbard band (UHB) and the lower Hubbard band (LHB) of the Cu $3d_{x^2-y^2}$ orbital, calculations based on the local density approximation (LDA) combined with the dynamical mean-field theory (DMFT) revealed the possibility of the closing of the so-called charge-transfer (CT) gap between UHB of the Cu $3d_{x^2-y^2}$ orbital and the O $2p$ band observed in the hole-doped cuprates~\cite{Das,Weber}.
That is, both metallic and SC states were suggested to appear by eliminating the AF order in the parent compounds of the T'-cuprates. 
The recent calculation of a two-particle self-consistent analysis using the three-band model~\cite{Ogura} revealed a monotonic decrease in $T_{\rm c}$ with electron doping, which is consistent with the experimental observation shown in Figure~\ref{figure1}b~\cite{Matsumoto}.

The undoped (Ce-free) superconductivity challenges the long-thought understanding that the parent compounds of the high-$T_{\rm c}$ cuprates are AF Mott insulators. 
To understand the mechanism of the undoped superconductivity, it is extremely important to clarify the evolution of the electronic state through the reduction process, for which the detailed investigation using single crystals is necessary.
One of critical questions is whether or not the undoped superconductivity in the T'-cuprates appears based on the strong electron correlation.
To answer this issue, it is significant to investigate the Cu-spin correlation, because localized Cu spins in LHB of the Cu $3d_{x^2-y^2}$ orbital are generated and expected to correlate with one another in the case of the strong electron correlation.
To attain these aims, we used T' single crystals of Ce-underdoped Pr$_{1.3-x}$La$_{0.7}$Ce$_x$CuO$_{4 + \delta}$ with $x=0.10$ and undoped (Ce-free) SC samples of La$_{1.8}$Eu$_{0.2}$CuO$_{4 + \delta}$.
In the next section, we start from the preparation of single crystals and the improved reduction annealing in Pr$_{1.3-x}$La$_{0.7}$Ce$_x$CuO$_{4 + \delta}$.

%%%%%%%%%%%%%%%%%%%%%%%%%%%%%%%%%%%%%%%%%%
\section{Evolution of the Electronic State through the Reduction Annealing}

Single crystals of Ce-underdoped Pr$_{1.3-x}$La$_{0.7}$Ce$_x$CuO$_{4+\delta}$ with $x=0.10$ were grown by the traveling-solvent floating-zone method~\cite{Risdi-PLCCO,Fujita-PD}.
In order to reduce excess oxygen in the as-grown crystals, we grew single crystals in air atmosphere, which is at a lower partial pressure of oxygen than formerly reported~\cite{Fujita-PD,Onose}.
The composition of as-grown crystals was determined by the inductively-coupled plasma (ICP) analysis to be Pr$_{1.17(1)}$La$_{0.73(1)}$Ce$_{0.10(1)}$Cu$_{1.00(1)}$O$_4$, which is close to the nominal composition. 
For the as-grown crystals, we performed the reduction annealing in vacuum of $\sim 10^{-4}$ Pa for 24 h at various temperatures.
In order to protect the surface of crystals during the annealing, the as-grown crystals were covered with polycrystalline powders having the same composition (protect annealing). 
This is an improved technique of the Brinkmann's~\cite{Brinkmann}.
From the iodometric titration, the oxygen content was confirmed to be reduced by 0.03(1) through the reduction annealing for Pr$_{1.3-x}$La$_{0.7}$Ce$_x$CuO$_{4+\delta}$ with $x=0.10$.
Moreover, through the reduction annealing, the $c$-axis lengths estimated from the X-ray diffraction measurements were reduced by 0.028(1) {\AA} for Pr$_{1.3-x}$La$_{0.7}$Ce$_x$CuO$_{4+\delta}$ with $x=0.10$, suggesting the removal of excess oxygen from the as-grown samples.
The scanning electron microscope also revealed that, in the reduced SC crystal, neither Cu metals nor rare-earth oxides were observed. 
Therefore, both the absence of Cu deficiency within the error of ICP and the oxygen reduction through the annealing suggest that the Cu deficiency model is not applicable to the present samples.
Polycrystalline samples of undoped (Ce-free) La$_{1.8}$Eu$_{0.2}$CuO$_{4+\delta}$ were prepared by a low-temperature technique using CaH$_2$ as a reductant and the subsequent annealing.
The SC samples were obtained through the reduction annealing at 700~$^{\circ}$C for 24 h in vacuum.
Details have been described in the literature~\cite{Takamatsu}.

\begin{figure}[tbp]
\centering
\includegraphics[width=1.0\linewidth]{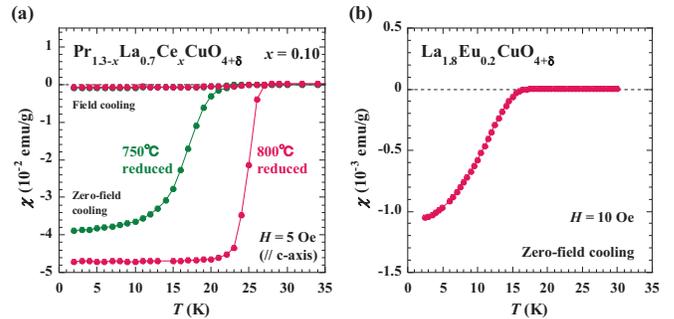}
\caption{Temperature dependence of the magnetic susceptibility of (\textbf{a}) Ce-underdoped Pr$_{1.3-x}$La$_{0.7}$Ce$_x$CuO$_{4+\delta}$ with $x=0.10$ and (\textbf{b}) undoped (Ce-free) La$_{1.8}$Eu$_{0.2}$CuO$_{4+\delta}$.} 
\label{figure2}
%\end{center}
\end{figure} 

Figure \ref{figure2} shows the temperature dependence of the magnetic susceptibility of Pr$_{1.3-x}$La$_{0.7}$Ce$_x$CuO$_{4+\delta}$ with $x=0.10$ and La$_{1.8}$Eu$_{0.2}$CuO$_{4+\delta}$.
For Pr$_{1.3-x}$La$_{0.7}$Ce$_x$CuO$_{4+\delta}$, the~Meissner diamagnetism appears in the 750 $^{\circ}$C-reduced crystal, and the relatively sharp SC transition with the onset of $~27$ K is observed in the 800 $^{\circ}$C-reduced crystal.
Formerly, Sun et al. reported from transport properties of reduced single crystals of Pr$_{1.3-x}$La$_{0.7}$Ce$_x$CuO$_{4+\delta}$ through the annealing in Ar that the crystals with $x \le 0.10$ are non-SC~\cite{Sun}.
The present successful observation of bulk superconductivity in Pr$_{1.3-x}$La$_{0.7}$Ce$_x$CuO$_{4+\delta}$ with $x=0.10$ suggests that the excess oxygen is effectively removed from the as-grown crystal through the protect annealing in vacuum at lower temperatures than in the conventional annealing.
The La$_{1.8}$Eu$_{0.2}$CuO$_{4+\delta}$ sample also shows the bulk Meissner diamagnetism below $\sim 15$ K, which is almost the same as that formerly reported~\cite{Takamatsu}.

\begin{figure}[tbp]
\centering
\includegraphics[width=1.0\linewidth]{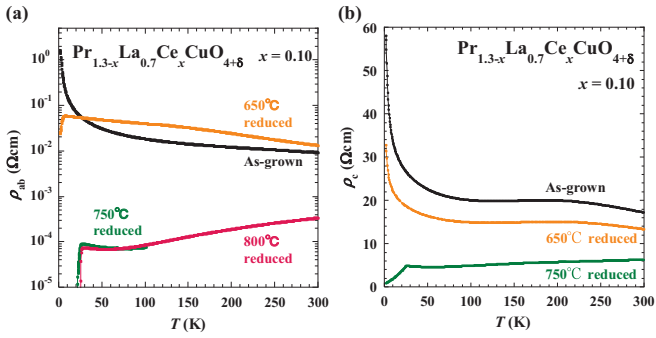} 
\caption{Temperature dependence of the (\textbf{a}) $ab$-plane and (\textbf{b}) $c$-axis electrical resistivity of Ce-underdoped Pr$_{1.3-x}$La$_{0.7}$Ce$_x$CuO$_{4+\delta}$ with $x=0.10$ in various annealing conditions.} \label{figure3}
\end{figure} 

The temperature dependence of $\rho_{\rm ab}$ in Pr$_{1.3-x}$La$_{0.7}$Ce$_x$CuO$_{4+\delta}$ with $x=0.10$ exhibits characteristic behaviors depending on the reduction condition, as shown in Figure \ref{figure3}a.
The as-grown crystal shows semiconducting temperature dependence and no trace of superconductivity.
For the 650 $^{\circ}$C-reduced crystal, $\rho_{\rm ab}$ is semiconducting at high temperatures, while a drop of $\rho_{\rm ab}$ is observed below $\sim 7$ K due to the SC transition. 
The drastic decrease in $\rho_{\rm ab}$ is observed in the 750 $^{\circ}$C- and 800 $^{\circ}$C-reduced crystals in which the bulk superconductivity appears.
Both crystals show a metallic behavior at high temperatures and an upturn at low temperatures, followed by the SC transition.
In~the normal state below $\sim 50$ K, $\rho_{\rm ab}$ exhibits log $T$ dependence.
This behavior, as well as the saturation of $\rho_{\rm ab}$ in the magnetic field and negative magnetoresistance~\cite{Adachi-JPSJ} is most likely due to the Kondo effect~\cite{Sekitani}.
The present evolution of $\rho_{\rm ab}$ through the reduction annealing suggests that the strongly localized state of carriers in the as-grown crystal changes to a metallic state with the Kondo effect in Pr$_{1.3-x}$La$_{0.7}$Ce$_x$CuO$_{4+\delta}$ with $x=0.10$.

Figure \ref{figure3}b shows the temperature dependence of the $c$-axis resistivity $\rho_{\rm c}$ for Pr$_{1.3-x}$La$_{0.7}$Ce$_x$CuO$_{4+\delta}$ with $x=0.10$.
For the as-grown and 650 $^{\circ}$C-reduced crystals, a hump is observed at $\sim$ 200 K, which is similar to that observed in Nd$_{2-x}$Ce$_x$CuO$_4$ with $x<0.14$~\cite{Onose} and Pr$_{1.3-x}$La$_{0.7}$Ce$_x$CuO$_{4+\delta}$ with $x \le 0.03$~\cite{Sun} and due to the opening of an AF pseudogap.
For the 750 $^{\circ}$C-reduced crystal, on the contrary, a simple metallic behavior without hump is observed, which resembles the behavior of $\rho_{\rm c}$ in SC Nd$_{2-x}$Ce$_x$CuO$_4$ with $x=0.15$~\cite{Onose}. 
Therefore, it is inferred that the AF pseudogap disappears in the 750 $^{\circ}$C-reduced SC crystal.

\begin{figure}[tbp]
\centering
\includegraphics[width=1.0\linewidth]{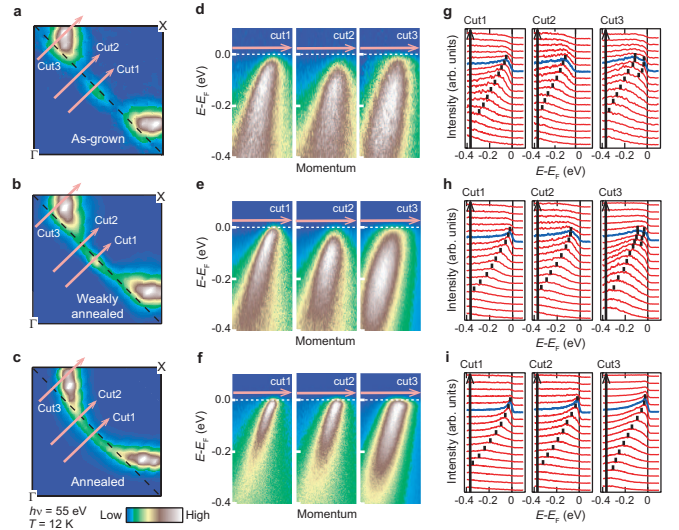}
\caption{(\textbf{a}--\textbf{c}) Fermi-surface mappings of as-grown, 650 $^{\circ}$C-reduced and 800 $^{\circ}$C-reduced crystals of Ce-underdoped Pr$_{1.3-x}$La$_{0.7}$Ce$_x$CuO$_{4+\delta}$ with $x = 0.10$, respectively. The intensities at the hot spots, the crossing points of the Fermi surface with the AF Brillouin zone boundary, are suppressed in the as-grown and 650 $^{\circ}$C-reduced crystals, while those are fully recovered in the 800 $^{\circ}$C-reduced crystal. (\textbf{d}--\textbf{f}) Intensity plots in the energy-momentum space for each crystal along each cut indicated in ({a}--{c}). (\textbf{g}--\textbf{i}) Energy-distribution curves plotted for each cut. Blue~energy-distribution curves are taken at positions of the Fermi wavenumber. Peak positions are marked by vertical bars. Quoted from~\cite{Horio}.} \label{figure4}
\end{figure} 

The electronic state of Pr$_{1.3-x}$La$_{0.7}$Ce$_x$CuO$_{4+\delta}$ was further investigated by the angle-resolved photoemission spectroscopy (ARPES) by Horio et al. using our as-grown and reduced single crystals, as shown in Figure \ref{figure4}~\cite{Horio}.
For the as-grown and 650 $^{\circ}$C-reduced crystals, the AF pseudogap is clearly observed around the hot spots, namely, crossing points of the Fermi surface with the AF Brillouin zone boundary.
For the 800 $^{\circ}$C-reduced crystal, on the contrary, the sharp quasi-particle peaks are observed on the entire Fermi surface without the signature of the AF pseudogap unlike the previous works~\cite{Armitage-PRL,Matsui}.
These results are consistent with the $\rho_{\rm c}$ results in Figure \ref{figure3}b and indicate the dramatic reduction of the AF correlation length and/or of magnetic moments in the 800 $^{\circ}$C-reduced crystal.
The quasi-particle peaks around the anti-nodal region are broader than those around the nodal region, suggesting the strong scattering of quasiparticles by the AF fluctuations and/or charge fluctuations in relation to the recent observation of the charge order in Nd$_{2-x}$Ce$_x$CuO$_4$~\cite{Eduardo}.

The Hall effect of the electron-doped T'-cuprates is highly informative.
Thin-film studies in Pr$_{2-x}$Ce$_x$CuO$_4$ reduced in the conventional annealing revealed that the Hall coefficient $R_{\rm H}$ was negative in the underdoped regime and increased with increasing $x$, followed by the sign change of $R_{\rm H}$ at $x \sim 0.17$~\cite{Dagan}.
Moreover, the Hall resistivity $\rho_{xy}$ exhibited a nonlinear behavior in the magnetic field in Pr$_{2-x}$Ce$_x$CuO$_4$ thin films~\cite{Li}.
Our results of $\rho_{xy}$ in Pr$_{1.3-x}$La$_{0.7}$Ce$_x$CuO$_{4+\delta}$ with $x=0.10$ also revealed a nonlinear behavior in the magnetic field.
These suggest that multiple carriers (most likely electrons and holes) reside in the T'-cuprates.
The existence of multiple carriers of electrons and holes has also been suggested from NMR in single crystals of Nd$_{2-x}$Ce$_x$CuO$_4$ and Pr$_{2-x}$Ce$_x$CuO$_4$~\cite{Jurkutat}.
Note that the undoped SC thin films exhibit positive $R_{\rm H}$ in the ground state, suggesting the existence of predominant hole-like carriers~\cite{Krockenberger-SciRep}.

%%%%%%%%%%%%%%%%%%%%%%%%%%%%%%%%%%%%%%%%%%
\section{Relationship between Cu-Spin Correlation and Superconductivity}

For the detailed investigation of the Cu-spin correlation, we used the $\mu$SR technique which is highly sensitive to the local magnetism in a sample and has an advantage to distinguish between a long-range and short-range magnetic order.
Zero-field (ZF) and longitudinal-field (LF) $\mu$SR measurements of Pr$_{1.3-x}$La$_{0.7}$Ce$_x$CuO$_{4+\delta}$ with $x=0.10$ and La$_{1.8}$Eu$_{0.2}$CuO$_{4+\delta}$ were performed using a pulsed positive muon beam at the Material and Life Science Experimental Facility at J-PARC in Japan and using a continuous positive muon beam at the Paul-Scherrer Institute (PSI) in Switzerland,~respectively.

\begin{figure*}[tbp]
\centering
\includegraphics[width=1.0\linewidth]{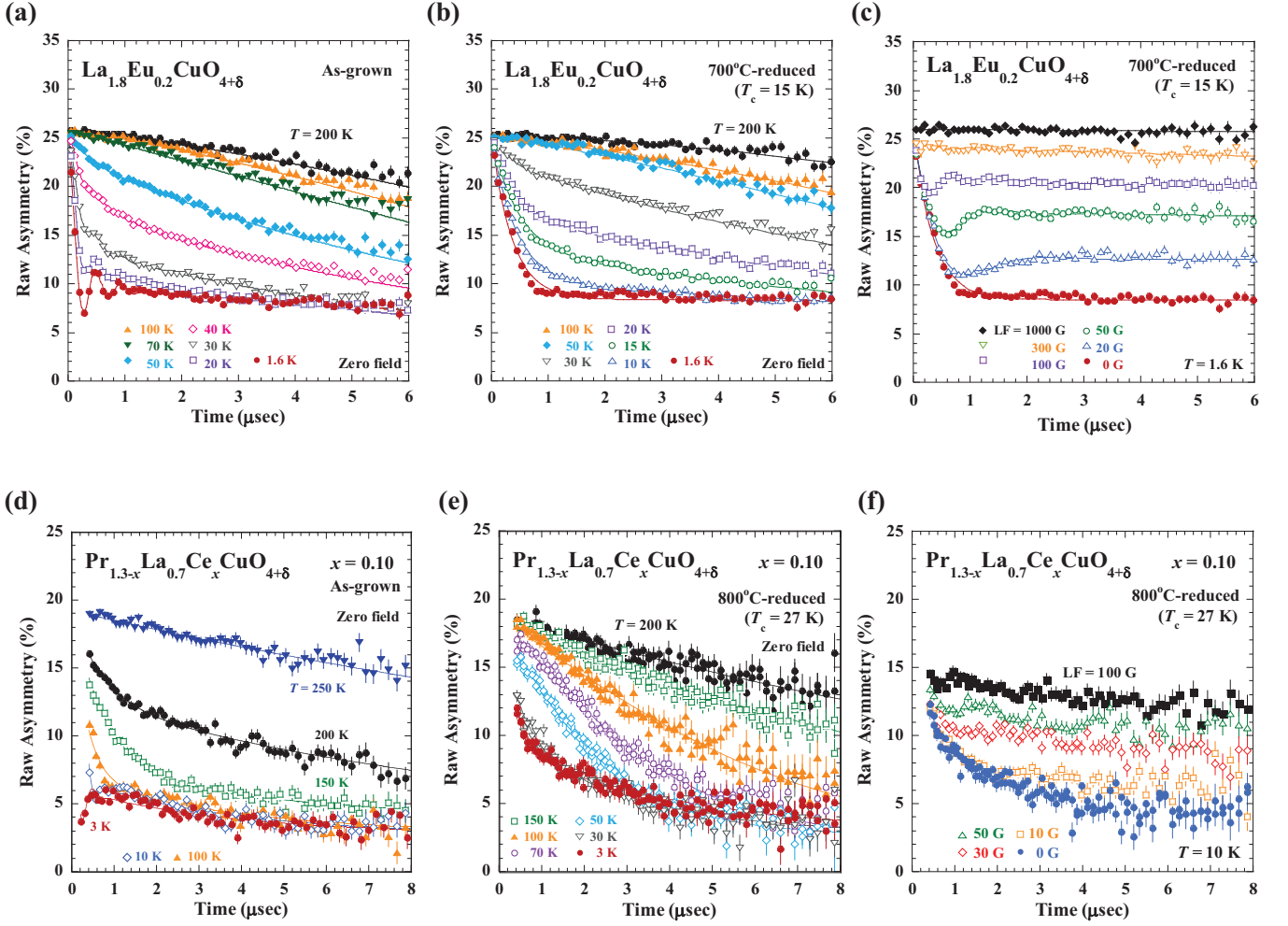}
\caption{{Zero-field (ZF) and longitudinal-field (LF)} %we noticed that some figures are not properly quoted，Please Provide Permission to reproduce,  you can click the likn to to apply for the copyright  permission: http://journals.jps.jp/page/permission
$\mu$SR time spectra of undoped (Ce-free) and Ce-underdoped T'-cuprates~\cite{Adachi-JPSJ2}. (\textbf{a},\textbf{b}) ZF-$\mu$SR time spectra of as-grown non-SC and 700 $^{\circ}$C-reduced SC samples of undoped (Ce-free) La$_{1.8}$Eu$_{0.2}$CuO$_{4+\delta}$; (\textbf{c}) LF-$\mu$SR time spectra at 1.6 K of 700 $^{\circ}$C-reduced SC samples of undoped La$_{1.8}$Eu$_{0.2}$CuO$_{4+\delta}$; (\textbf{d},\textbf{e}) ZF-$\mu$SR time spectra of as-grown non-SC and 800~$^{\circ}$C-reduced SC crystals of Ce-underdoped Pr$_{1.3-x}$La$_{0.7}$Ce$_x$CuO$_{4+\delta}$ with $x=0.10$; (\textbf{f}) LF-$\mu$SR time spectra at 10 K of 800 $^{\circ}$C-reduced SC crystal of Pr$_{1.3-x}$La$_{0.7}$Ce$_x$CuO$_{4+\delta}$ with $x=0.10$. Solid lines in ({a})--({e}) indicate the best-fit results using the analysis function described in the literature.} \label{figure5}
\end{figure*}

Figure \ref{figure5}a,b show the ZF-$\mu$SR time spectra of as-grown non-SC and 700 $^{\circ}$C-reduced SC samples of La$_{1.8}$Eu$_{0.2}$CuO$_{4+\delta}$, respectively~\cite{Adachi-JPSJ2}.
For both samples, Gaussian-like slow depolarizations due to nuclear dipole fields are observed at high temperatures, indicating a paramagnetic state of Cu spins.
With decreasing temperature, fast depolarizations of muon spins are observed, suggesting the development of the Cu-spin correlation.
For the as-grown sample, clear muon-spin precessions appear at low temperatures, indicating the formation of a long-range magnetic order.
For the reduced SC sample, the spectrum consists of a fast depolarization without precession and a following time-independent behavior above 1 $\mu$s, which is typical of a short-range magnetically ordered state.
In~the LF-$\mu$SR spectra at 1.6 K shown in Figure \ref{figure5}c~\cite{Adachi-JPSJ2}, a parallel shift up with LF indicates the formation of a static magnetically ordered state in ZF.
In Pr$_{1.3-x}$La$_{0.7}$Ce$_x$CuO$_{4+\delta}$ with $x=0.10$, the as-grown non-SC crystal exhibits muon-spin precessions at low temperatures as shown in Figure~\ref{figure5}d~\cite{Adachi-JPSJ2}. 
The spectra of the 800 $^{\circ}$C-reduced SC crystal in Figure \ref{figure5}e~\cite{Adachi-JPSJ2} show both fast and slow depolarizations without muon-spin precession at 3 K.
The LF-$\mu$SR spectra at 10 K shown in Figure~\ref{figure5}f~\cite{Adachi-JPSJ2} reveal a parallel shift as well as a slow depolarization with LF, suggesting the existence of both a short-range magnetic order and fluctuating Cu spins.
These results indicate the coexisting state of the superconductivity and the short-range magnetic order in both La$_{1.8}$Eu$_{0.2}$CuO$_{4+\delta}$ and Pr$_{1.3-x}$La$_{0.7}$Ce$_x$CuO$_{4+\delta}$ with $x=0.10$.
Note that $\mu$SR results of the undoped (Ce-free) thin film of La$_{2-x}$Y$_x$CuO$_{4+\delta}$ have shown the development of the Cu-spin correlation at low temperatures~\cite{Kojima}.

The analysis of volume fractions of SC and short-range magnetically ordered regions leads to the detailed information on the electronic state in the T'-cuprates.
The magnetic volume fractions were estimated from the $\mu$SR results~\cite{Adachi-JPSJ2}.
For the reduced SC La$_{1.8}$Eu$_{0.2}$CuO$_{4+\delta}$, the volume fraction of the short-range magnetically ordered region is almost 100\% in the ground state.
As the SC volume fraction is estimated from the Meissner fraction to be above 15\%, both the SC and short-range magnetically ordered regions coexist in the sample.
For the reduced SC Pr$_{1.3-x}$La$_{0.7}$Ce$_x$CuO$_{4+\delta}$ with $x=0.10$, the estimated magnetic volume fraction is 80\% in the ground state.
The SC volume fraction was estimated from the recovery of the electronic specific heat coefficient in the ground state by the application of magnetic field.
It was at least 60\%, suggesting the coexistence of the SC and short-range magnetically ordered regions in the sample.
Numerically, a part of the short-range magnetically ordered regions appear to be spatially overlapped with the SC regions. 
However, it is plausible that the short-range magnetically ordered and SC regions are phase-separated, because the volume fraction of the short-range magnetically ordered region might be overestimated due to electronic dipole fields extending to the paramagnetic SC region.

\begin{figure}[tbp] 
\centering
\includegraphics[width=1.0\linewidth]{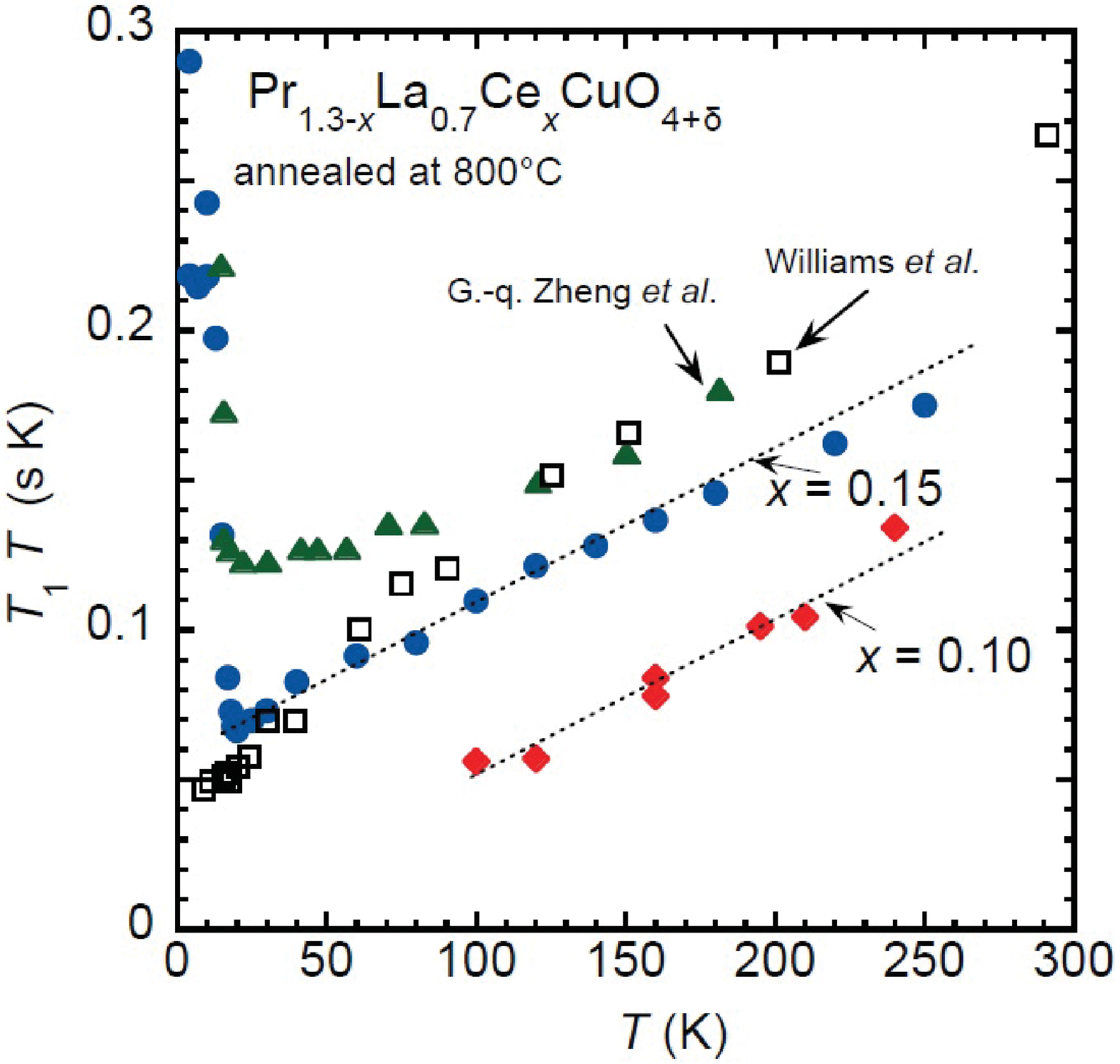}
\caption{Temperature dependence of %we noticed that some figures are not properly quoted，Please Provide Permission to reproduce
 $T_1T$ for reduced SC Pr$_{1.3-x}$La$_{0.7}$Ce$_x$CuO$_{4+\delta}$ with $x=0.10$ and 0.15. Quoted from~\cite{Yamamoto}.} \label{figure6}
%\end{center} 
\end{figure}

The Cu-spin dynamics was further investigated by NMR using our reduced SC T'-samples by Yamamoto et al.~\cite{Yamamoto} and Fukazawa~\cite{Fukazawa}.
As shown in Figure \ref{figure6}, the temperature dependence of $T_1 T$, where $1/T_1$ is the $^{63}$Cu nuclear spin-lattice relaxation rate exhibited the Curie-Weiss behavior in a~wide temperature range for Pr$_{1.3-x}$La$_{0.7}$Ce$_x$CuO$_{4+\delta}$ with $x=0.10$ and 0.15, suggesting the existence of strong AF fluctuations in the underdoped and overdoped T'-cuprates.
For the undoped (Ce-free) La$_{1.8}$Eu$_{0.2}$CuO$_{4+\delta}$, the bulk superconductivity was evidenced by the significant decrease in the Knight shift below $T_{\rm c}$ and AF fluctuations were observed in the normal state from the temperature dependence of $1/T_1$.
These results suggest that AF fluctuations are related to the formation of electron pairs in the T'-cuprates as well as in the hole-doped cuprates.

%%%%%%%%%%%%%%%%%%%%%%%%%%%%%%%%%%%%%%%%%%
\section{Superconducting Pairing Symmetry}

\begin{figure}[tbp]
\centering
\includegraphics[width=1.0\linewidth]{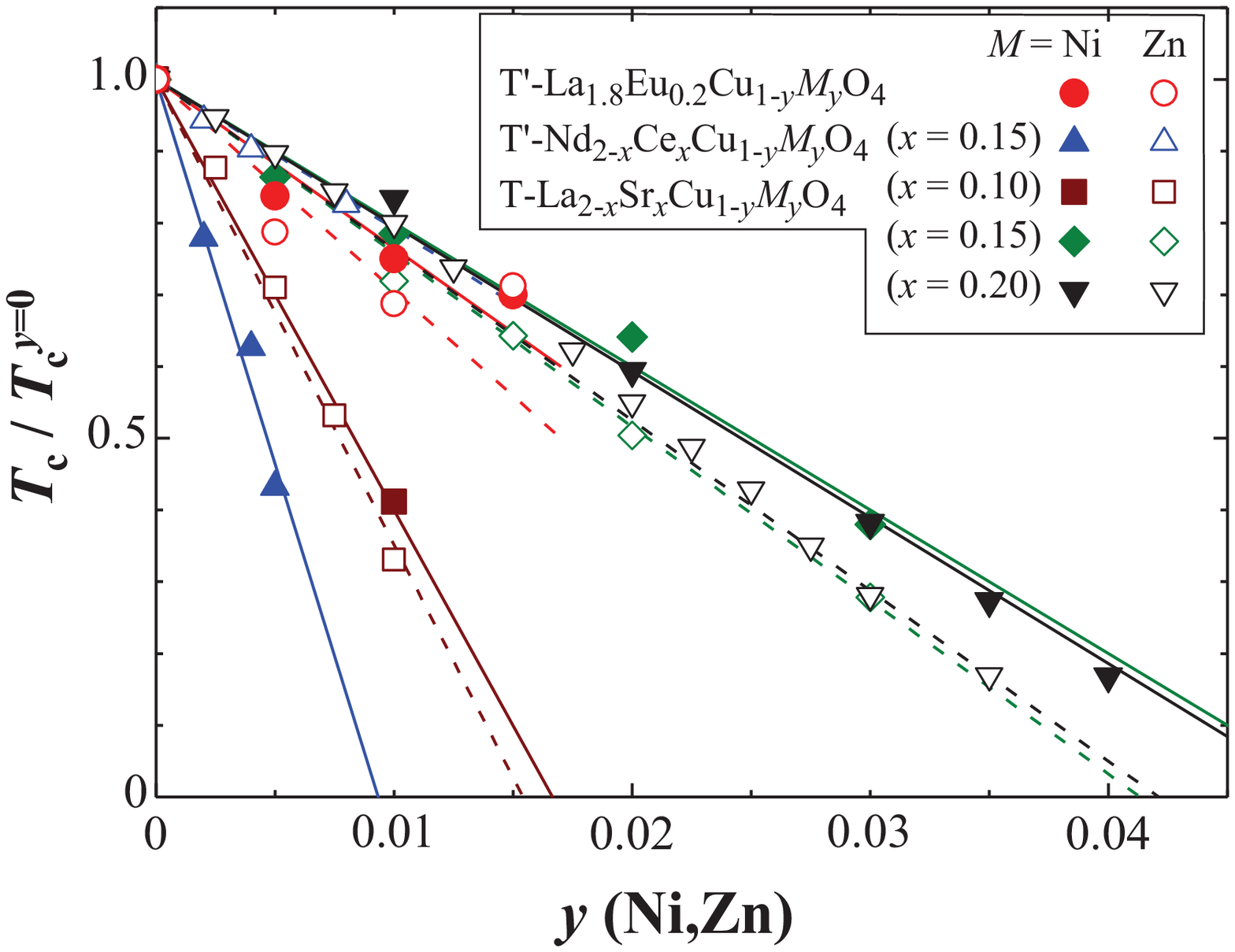}
\caption{Impurity-concentration dependence of $T_{\rm c}$ normalized by $T_{\rm c}$ at $y=0$ $T_{\rm c}^{y=0}$ for undoped (Ce-free) La$_{1.8}$Eu$_{0.2}$Cu$_{1-y}$(Zn,Ni)$_y$O$_{4+\delta}$ together with the preceding results of Nd$_{2-x}$Ce$_x$Cu$_{1-y}$(Zn,Ni)$_y$O$_{4+\delta}$ ($x=0.15$)~\cite{Tarascon} and of hole-doped La$_{2-x}$Sr$_x$Cu$_{1-y}$(Zn,Ni)$_y$O$_4$ \mbox{($x=0.10, 0.15, 0.20$)}~\cite{Tarascon,Sreedhar,Uchida}. Filled and open symbols indicate data of the Ni and Zn substitution, respectively. Solid and dashed lines are guides to the eye.}
\label{figure7}
\end{figure}

While the SC pairing symmetry in the hole-doped cuprates is widely recognized to be $d$-wave, it is controversial in the electron-doped T'-cuprates.
Early studies of impurity effects in Nd$_{2-x}$Ce$_x$Cu$_{1-y}$(Zn,Ni)$_y$O$_4$ with $x=0.15$ revealed that $T_{\rm c}$ was depressed by the magnetic impurity Ni much more rapidly than by the nonmagnetic impurity Zn~\cite{Tarascon}.
This is typical of conventional $s$-wave superconductors.
On the contrary, ARPES~\cite{Sato,Matsui} and penetration depth~\cite{Kokales} experiments etc. insisted the occurrence of $d$-wave superconductivity in Nd$_{2-x}$Ce$_x$CuO$_{4+\delta}$ with $x=0.15$ and Pr$_{2-x}$Ce$_x$CuO$_{4+\delta}$ with $x=0.15$.
To clarify the SC pairing symmetry of the undoped (Ce-free) superconductivity in the T'-cuprates, impurity effects on $T_{\rm c}$ were precisely investigated in La$_{1.8}$Eu$_{0.2}$Cu$_{1-y}$(Zn,Ni)$_y$O$_{4+\delta}$~\cite{Ohashi}.
Figure \ref{figure7} displays the impurity-concentration dependence of $T_{\rm c}$ normalized by $T_{\rm c}$ of $y=0$ $T_{\rm c}^{y=0}$ in La$_{1.8}$Eu$_{0.2}$Cu$_{1-y}$(Zn,Ni)$_y$O$_{4+\delta}$ together with the preceding results of T'-cuprates~\cite{Tarascon} and hole-doped La$_{2-x}$Sr$_x$Cu$_{1-y}$(Zn,Ni)$_y$O$_4$~\cite{Tarascon,Sreedhar,Uchida}.
Obviously, the depression of $T_{\rm c}$ is almost the same between the Zn and Ni substitution in contrast to the preceding results of Nd$_{2-x}$Ce$_x$Cu$_{1-y}$(Zn,Ni)$_y$O$_{4+\delta}$ with $x=0.15$~\cite{Tarascon}.
Moreover, the change of $T_{\rm c}/T_{\rm c}^{y=0}$ in La$_{1.8}$Eu$_{0.2}$Cu$_{1-y}$(Zn,Ni)$_y$O$_{4+\delta}$ follows the results of La$_{2-x}$Sr$_x$Cu$_{1-y}$(Zn,Ni)$_y$O$_4$ in the optimally doped $x=0.15$ and overdoped $x=0.20$. 
Therefore, in La$_{1.8}$Eu$_{0.2}$CuO$_{4+\delta}$, the depression of $T_{\rm c}$ by the Zn and Ni substitution is probably due to the pair-breaking effect and the SC pairing symmetry is $d$-wave mediated by spin fluctuations as in the case of the hole-doped cuprates.
In fact, our $\mu$SR~\cite{Adachi-JPSJ2} and NMR~\cite{Fukazawa} results also revealed strong spin fluctuations in La$_{1.8}$Eu$_{0.2}$CuO$_{4+\delta}$.
The characteristic temperature dependence of $1/T_1$ observed in La$_{1.8}$Eu$_{0.2}$CuO$_{4+\delta}$ implies the existence of nodal lines in the SC gap, supporting the above statement of the $d$-wave superconductivity~\cite{Fukazawa}.
Moreover, optical studies in the SC thin film of Pr$_2$CuO$_x$ with $x \simeq 4$ have suggested that the temperature dependence of the magnetic penetration depth $\lambda$ exhibits the $d$-wave-like behavior~\cite{Chanda}.
In future, we will further investigate the SC pairing symmetry by the transverse-field $\mu$SR technique.
Accumulating results of $\lambda$, inversely proportional to the SC carrier density, in electron-doped T'-cuprates~\cite{Kojima,Homes,Nugroho,Homes2} and infinite-layer cuprates~\cite{Satoh} revealed that the $T_{\rm c}$ vs. $\lambda$ relation seemed not to intersect the origin, which is contrary to the results of hole-doped cuprates suggesting the Bose--Einstein-condensation-like pairing mechanism~\cite{Uemura}.

%*****************************************************************************************
\section{Electronic Structure Model Based on the Strong Electron Correlation}
%*****************************************************************************************

Our transport and $\mu$SR results of Pr$_{1.3-x}$La$_{0.7}$Ce$_x$CuO$_{4+\delta}$ with $x=0.10$ and La$_{1.8}$Eu$_{0.2}$CuO$_{4+\delta}$ revealed the following evolution of the electronic state through the reduction annealing.
The $\rho_{\rm ab}$ showed that a strongly localized state of carriers in the as-grown sample changes to a metallic state with the Kondo effect in the reduced SC sample. 
The Hall resistivity of the reduced SC sample revealed the existence of multiple carriers. 
The $\mu$SR spectra revealed that, in the ground state, a long-range magnetic order of Cu spins in the as-grown sample changed to a short-range one coexisting with the superconductivity in the reduced SC samples. 
It would be a reasonable speculation that an ideal T'-cuprate in which the excess oxygen is completely removed exhibits no AF order.
The formation of the short-range magnetic order due to a very small amount of excess oxygen in the reduced SC sample suggests that the Cu spins are correlated with one another in the absence of excess oxygen in the ideal T'-cuprates. 
Therefore, the T'-cuprates exhibiting the undoped (Ce-free) superconductivity are regarded as strongly correlated electron systems as well as the hole-doped SC cuprates. 
Not the simple band-metal state without the electron correlation~\cite{Massidda} but both the doublon-holon model~\cite{Yokoyama}, calculations using LDA with DMFT~\cite{Das,Weber} and the two-particle self-consistent analysis~\cite{Ogura} under the electron correlation would be able to explain the superconductivity in the undoped (Ce-free) and Ce-underdoped T'-cuprates.

\begin{figure*}[tbp]
\centering
\includegraphics[width=0.8\linewidth]{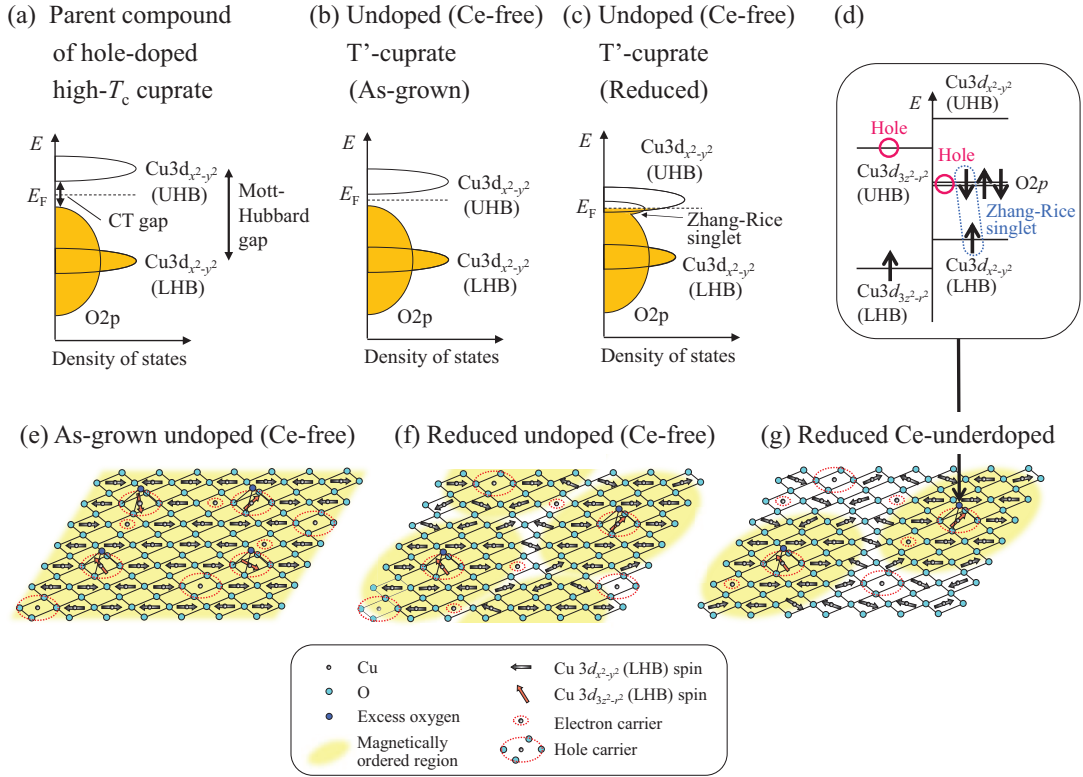}
\caption{(\textbf{a}--\textbf{c}) Electronic structure models for ({a}) the parent compound of hole-doped high-$T_{\rm c}$ cuprate; (b) the as-grown undoped (Ce-free) T'-cuprate and ({c}) the reduced SC undoped (Ce-free) T'-cuprate. Arrows in ({a}) denote the charge-transfer (CT) and Mott-Hubbard gaps. (\textbf{d}) Electronic structure of the CuO$_2$ plane near the excess oxygen in undoped (Ce-free) T'-cuprates. (\textbf{e}--\textbf{g})~Schematic drawings of electron and hole carriers and Cu spins in the CuO$_2$ plane for ({e}) the as-grown non-SC undoped (Ce-free); ({f}) the reduced SC undoped (Ce-free) and ({g}) the reduced SC Ce-underdoped~T'-cuprates.} \label{figure8}
\end{figure*}

The present transport and $\mu$SR results are able to be understood in terms of the electronic structure model based on the strong electron correlation.
The electronic structure models and schematic drawings of Cu spins and carriers in the CuO$_2$ plane are displayed in Figure \ref{figure8}a--c,e--g, \mbox{respectively~\cite{Adachi-JPSJ,Adachi-JPSJ2}}.
In~the hole-doped high-$T_{\rm c}$ cuprates, the parent compounds are CT insulators as shown in Figure \ref{figure8}a characterized by the CT gap between UHB of the Cu $3d_{x^2-y^2}$ orbital and O $2p$ band.
In~the ideal undoped (Ce-free) T'-cuprates without excess oxygen shown in Figure \ref{figure8}c, on the other hand, the square-planar coordination of Cu and oxygen in the CuO$_2$ plane gives rise to a decrease in the Madelung energy of the Cu $3d$ orbitals compared with the octahedral coordination in the hole-doped cuprates. 
This brings about the lowering energy of UHB of the Cu $3d_{x^2-y^2}$ orbital, leading to the mixing of UHB with the Zhang-Rice singlet band.
That is, the CT gap is collapsed, so that both mobile electrons and holes simultaneously emerge at the Fermi level in the parent compounds of the T'-cuprates, leading to the appearance of superconductivity without nominal doping of electrons by the Ce substitution.
This is plausible, because the calculation of the Madelung energy results in a~decrease in the energy of the Cu $3d$ orbitals by $\sim 4$ eV compared with the hole-doped cuprates with the CT gap $\sim 2$ eV~\cite{Madelung}. 
In the case of excess oxygen residing adjacent to the CuO$_2$ plane, assuming that two holes are produced in the CuO$_2$ plane by one excess oxygen, the doped holes in the CuO$_2$ plane tend to be localized near the excess oxygen due to the disorder of the electrostatic potential.
The holes are doped into the Cu $3d$ and O $2p$ orbitals taking into account the strong electron correlation. 
This is because the energy of the Cu $3d$ orbitals is raised by the excess oxygen due to increasing coordination number, as shown in Figure \ref{figure8}d.
The doped hole in the Cu $3d$ orbital resides not in LHB of the Cu $3d_{x^2-y^2}$ orbital but in UHB of the Cu $3d_{3z^2-r^2}$ orbital in order to gain the energy of the Zhang--Rice singlet of Cu $3d_{x^2-y^2}$ (LHB) and O $2p$ spins.
This is reasonable, because the splitting of Cu $3d_{x^2-y^2}$ and Cu $3d_{3z^2-r^2}$ in energy is smaller than the Mott--Hubbard gap of $\sim 8$ eV.
Accordingly, Cu $3d_{3z^2-r^2}$ (LHB) free spins are probably induced at the Cu site adjacent to the excess oxygen.
Both hole and electron carriers tend to be localized because of the disorder of the electrostatic potential near the excess oxygen, resulting in the recovery of the AF order of Cu $3d_{x^2-y^2}$ (LHB) spins.
This is able to explain the increase in $\rho_{\rm ab}$ due to the strong localization and the hump of the temperature dependence of $\rho_{\rm c}$ due to the opening of the AF pseudogap.

In the CuO$_2$ plane in the as-grown non-SC samples of La$_{1.8}$Eu$_{0.2}$CuO$_{4+\delta}$ and Pr$_{1.3-x}$La$_{0.7}$Ce$_x$CuO$_{4+\delta}$ with $x=0.10$, because of the moderate number of excess oxygen, carriers are strongly localized, as confirmed by $\rho_{\rm ab}$~\cite{Adachi-JPSJ}, and the AF order is long-ranged, as illustrated in Figure \ref{figure8}e. 
In reduced SC La$_{1.8}$Eu$_{0.2}$CuO$_{4+\delta}$, as shown in Figure \ref{figure8}f, the CuO$_2$ plane should be mostly covered by short-range magnetically ordered regions, and superconductivity should appear around the boundary between the short-range magnetically ordered regions.
In reduced SC Pr$_{1.3-x}$La$_{0.7}$Ce$_x$CuO$_{4+\delta}$ with $x=0.10$, because the amount of excess oxygen is smaller than that of reduced SC La$_{1.8}$Eu$_{0.2}$CuO$_{4+\delta}$, the short-range magnetically ordered regions decrease in correspondence to the increase in the SC volume fraction, as shown in Figure \ref{figure8}g.
It is noted that the introduction of excess oxygen gives rise to free Cu spins of LHB of the Cu $3d_{3z^2-r^2}$ orbital in the CuO$_2$ plane just around itself, leading to the occurrence of the Kondo effect as confirmed by $\rho_{\rm ab}$ of reduced Pr$_{1.3-x}$La$_{0.7}$Ce$_x$CuO$_{4+\delta}$ with $x=0.10$~\cite{Adachi-JPSJ}.
Accordingly, it is suggested that the short-range magnetically ordered regions are formed around excess oxygen and superconductivity appears far from the excess oxygen. 

As mentioned above, the ARPES experiment of Pr$_{1.3-x}$La$_{0.7}$Ce$_x$CuO$_{4+\delta}$ with $x=0.10$ by \mbox{Horio et al.~\cite{Horio}} revealed that the AF pseudogap was closed on the whole Fermi surface, which is contrary to the former ARPES results in the T'-cuprates~\cite{Armitage-PRL,Matsui}.
Considering both results of ARPES without the AF pseudogap and $\mu$SR with the short-range magnetic order, it is likely that Cu spins forming the short-range magnetic order have tiny magnetic moments. 
That is, the reduction annealing gives rise to not only the change of the long-range AF order to the short-range one but also the reduction of the magnetic moments of Cu spins, which is compatible with the decrease in the internal magnetic field at the muon site estimated from $\mu$SR results~\cite{Adachi-JPSJ2}.

%*****************************************************************************************
\section{Future Issues}
%*****************************************************************************************

The present transport and $\mu$SR results of the undoped (Ce-free) and Ce-underdoped T'-cuprates uncovered that the superconductivity appeared under the strong electron correlation as in the case of the hole-doped cuprates.
Here we discuss remaining unsolved issues of the T'-cuprates.

One is that the oxygen dynamics through the reduction annealing must be directly clarified.
To date, several candidates of effects of the reduction annealing have been proposed; the removal of excess oxygen at the apical site~\cite{Radaelli,Schultz}, filling up Cu deficiencies~\cite{Mang,Kang}, etc.
Our crystal of Pr$_{1.3-x}$La$_{0.7}$Ce$_x$CuO$_{4+\delta}$ with $x=0.10$ seems to be the case of the reduction of excess oxygen.
Intriguing in our results is that the introduction of Cu vacancies into SC La$_{1.8}$Eu$_{0.2}$Cu$_{1-y}$O$_{4+\delta}$ does not change $T_{\rm c}$ so much up to $y=0.015$~\cite{Ohashi}, which seems to be incompatible with the Cu-deficiency model.
In~any case, this issue would be clarified by the recent high resolution neutron and X-ray diffraction experiments. 
Another issue is about carriers in the undoped (Ce-free) SC T'-cuprates.
Our proposed electronic structure model indicates the presence of both electron and hole carriers, which is compatible with the present and former~\cite{Li} Hall resistivity and NMR~\cite{Jurkutat} results suggesting the existence of multiple carriers.
On~the other hand, the ARPES results shown in Figure \ref{figure4} in Pr$_{1.3-x}$La$_{0.7}$Ce$_x$CuO$_{4+\delta}$ with $x=0.10$ showed a~hole-like Fermi surface without the AF pseudogap~\cite{Horio}.
This is probably explained by considering that the overlapping UHB of the Cu $3d_{x^2-y^2}$ orbital and Zhang--Rice band form a mixing band in which both hole-like and electron-like carriers reside. 
The actual carrier concentration is also under debate.
It has been suggested from the Hall effect results~\cite{Gauthier}, etc. that the removal of excess oxygen does not simply correspond to the electron doping.
Intriguing is the estimation of the electron concentration from the Fermi-surface volume observed in ARPES that $T_{\rm c}$ is unchanged in a wide electron concentration range in Pr$_{1.3-x}$La$_{0.7}$Ce$_x$CuO$_{4+\delta}$~\cite{Horio}.
On the contrary, very recent ARPES results in Pr$_{1-x}$LaCe$_x$CuO$_{4+\delta}$ revealed that $T_{\rm c}$ exhibited a parabolic change against the electron concentration estimated from the Fermi-surface volume~\cite{Song}. 

The hole doping into the parent compounds of the T'-cuprates is an interesting way to clarify the electronic structure.
Based on the proposed electronic structure model, the parent compounds of the T'-cuprates are no longer reference materials and $T_{\rm c}$ might continue to increase with hole doping into the parent compounds.
Takamatsu et al. reported that the hole doping by the Sr/Ca substitution for La in La$_{1.8-x}$Eu$_{0.2}$(Ca,Sr)$_x$CuO$_{4+\delta}$ results in the decrease in $T_{\rm c}$~\cite{Takamatsu,Takamatsu-PP}.
As Zn/Ni impurity effects on $T_{\rm c}$ in La$_{1.8}$Eu$_{0.2}$Cu$_{1-y}$(Zn,Ni)$_y$O$_{4+\delta}$ suggest that the electronic state of the parent compounds is similar to the overdoped regime of the hole-doped cuprates~\cite{Ohashi}, it may be natural that further hole doping results in the decrease in $T_{\rm c}$. %missing the conclusions?

\vspace{6pt}

%%%%%%%%%%%%%%%%%%%%%%%%%%%%%%%%%%%%%%%%%%
\acknowledgments{Our works were done in collaboration with Yosuke Mori, Akira Takahashi, Takuya Konno, Taro Ohgi, Tomohisa Takamatsu, Kensuke M. Suzuki, Koki Ohashi, Malik Angelh Baqiya, Koshi Kurashima, Masatsune Kato, Isao Watanabe, Masanori Miyazaki, Akihiro Koda and Ryosuke Kadono. %please provide full name, the abbreviations is not allowed
Our works were supported by JSPS KAKENHI Grant Number 23540399 and 23108004 and by Sophia University Special Grant for Academic Research.}

%=====================================
% References, variant A: internal bibliography
%=====================================
\bibliographystyle{mdpi}
\renewcommand\bibname{References}

% The following MDPI journals use author-date citation: Arts, Econometrics, Economies, Genealogy, Humanities, IJFS, JRFM, Laws, Religions, Risks, Social Sciences. For those journals, please follow the formatting guidelines on http://www.mdpi.com/authors/references

%%%%%%%%%%%%%%%%%%%%%%%%%%%%%%%%%%%%%%%%%%
\end{document}